# Atom- photon entanglement beyond the multi-photon resonance condition

**Zeinab Kordi, Saeed Ghanbari, and Mohammad Mahmoudi**

Department of Physics, University of Zanjan, University Blvd, 45371-38791, Zanjan, Iran

**Abstract:** The density matrix equations of motion in near-degenerate three-level V-type closed-loop atomic system are calculated numerically in Floquet frame. The dynamical behavior of atom- photon entanglement between the dressed atom and its spontaneous emission is studied in semi classical approach beyond the two-photon resonance condition in such a system. The quantum entropy of these two subsystems is investigated by using the von Neumann entropy. It is shown that, the degree of entanglement measure (DEM) can be controlled via the intensity and the detuning of coupling optical field and quantum interference induced by spontaneous emission. Moreover in the absence of quantum interference the steady state behavior of DEM can be achieved even in beyond the two- photon resonance condition. Furthermore in the absence of quantum interference for special parameters of Rabi frequency and detuning of driving laser field disentanglement can be occurred. Also the electromagnetically induced transparency condition can be obtained when the system is disentangled.

## Introduction

Quantum entanglement is a profound concept in quantum mechanics. For first the time, Einstein, Podolsky and Rosen in their elegant paper [1] introduced this mysterious and bizarre characteristic of nature. "Einstein hoped to show that quantum theory could not describe certain intuitive "elements of reality" and thus was either incomplete". Schrödinger described this feature as "entangled" (*verschränkt*) for two electrons at their first measurement [2]. Entanglement has been appeared as a nonlocal property of nature. "Einstein argued that the probability wave could not act simultaneously at different places without violating the theory of relativity"[3]. So far physicists have shown all physics is compatible with Einstein's relativity and quantum mechanics is complete. Beyond the philosophical aspects of quantum entanglement we would like to work on many applications of this interesting concept.

Mathematically, the quantum state of an entangled system cannot be described as a tensor product of the quantum states of the subsystems [4]. In fact when two subsystems are entangled, the measuring on first one tells us about the information of second one. Working with a computer that operates according to quantum mechanics, will be a great achievement in science [5]. Entanglement plays an important role in quantum computer processing and quantum communication[6,7] quantum teleportation and entanglement swapping [8,9], quantum super dense coding [10], quantum error correction [11,12], quantum cryptography [13-15],entanglement distillation[16] and quantum computing [17- 25].

Quantum entanglement can be generated due to the interaction between different parts of a system consisting of atoms, photons or a mixture of atoms and photons. Much effort has been devoted to investigate the interaction between light and ensemble of ultracold atoms [26]. "The fragile nature of ultracold quantum ensembles limits one to dispersive light-matter interactions excluding near resonant excitations followed by spontaneous emission."[27-29]. Furthermore entanglement between light and atomic ensembles has been extensively investigated [30,31].

In this study we would like to focus on quantum interference. In fact quantum interference induced by spontaneous emission plays a key role to control the entanglement between the atomic ensemble and spontaneous emission. Quantum interference between coupled transitions has this potential to trapping of the population in one of the atomic excited levels, thereby eliminating the population in the other levels [32, 33].

The effect of quantum interference on optical bistability in the three-level V-type atomic system in two-photon resonance condition and beyond it has been studied[34,35]. Also Entanglement between atom and its spontaneous emission field in a $\Lambda$ type atomic system has been investigated [36, 37]. The effect of quantum interference on the behavior of entanglement between three- level V-type atomic system and its spontaneous emission has been studied in two-photon resonance condition [38, 39]. Furthermore, it has been illustrated that, in the absence of quantum interference the atom–photon entanglement in closed-loop quantum systems can be controlled via the relative phase of driving optical fields [40, 41]. Here we discuss the dynamical and stationery behavior of atom- photon entanglement in a closed loop three- level V-type atomic system beyond the two-photon resonance condition. The light propagation through closed-loop atomic media beyond the multi-photon resonance condition has been investigated [42]. We show that, the time evolution of DEM has the oscillatory behavior beyond the two-photon resonance condition. We used the Floquet decomposition [43] to solve the stationary solution of density matrix equations of motion and to investigate DEM behavior versus the detuning of coupling beam and probe beam. It is demonstrated that the steady state DEM can be controlled via intensity or detuning of the coupling fields. In fact by considering the Floquet method we could investigate the portion of coupling and probe beam to create entanglement between the atomic ensemble and its spontaneous emission. It is obvious that, the DEM is so sensitive to the Rabi frequency and detuning of the coupling field. Moreover quantum interference is another parameter which could control the DEM in such a system. The physics behind of atom-photon entanglement behavior can be understood via population distribution of the dressed states [38].

We have shown that, the DEM can be happened due to the strong coupling between light and matter. When atoms and photons are partially entangled the information of the position of the atoms among the levels is not achievable which means atom and photon not as separate entities While in electromagnetically induced transparency condition the system will be disentangled because of coherent population trapping [44-47]. Under disentanglement situation the atoms and the photons are separated and all of the atoms are in a dressed state as a dark state.

**Model and equations**

We consider a pump-probe V-type three-level atomic system, containing single ground state $|1\rangle$ and two closely excited states $|2\rangle$ and $|3\rangle$, which is driven by a coupling laser field and one probe field as depicted in fig. 1. This type of atomic system can be prepared in $D_2$ transitions of cesium vapor at room temperature [48]. The transition $|1\rangle - |2\rangle$ is excited by a probe laser field with frequency $\omega_p$ and the transition $|1\rangle - |3\rangle$ is excited by a coupling field with frequency $\omega_c$. The spontaneous emissions from the upper levels $|2\rangle$ and $|3\rangle$ to the ground level $|1\rangle$ are denoted by $2\gamma_2$ and $2\gamma_3$ respectively. The semi classical Hamiltonian in interaction picture under dipole and rotating-wave approximation is

$$H = -\hbar\Omega_p e^{i\Delta_p t}|2\rangle\langle 1| - \hbar\Omega_c e^{i\Delta_c t}|3\rangle\langle 1| + H.C., \tag{1}$$

where $\Omega_p = \vec{E}_p \cdot \vec{d}_{12}/\hbar$ and $\Omega_c = \vec{E}_c \cdot \vec{d}_{13}/\hbar$ are the Rabi frequencies for the probe and coupling fields respectively. $\vec{d}_{12}$ and $\vec{d}_{13}$ stand for dipole moment of two transitions. The detuning of the probe and coupling fields with respect to the corresponding transition frequencies are defined by $\Delta_p = \omega_{21} - \omega_p$ and $\Delta_c = \omega_{31} - \omega_c$, respectively. The total electromagnetic field can be described as

$$\vec{E} = \vec{E}_p \exp(-i\omega_p t) + \vec{E}_c \exp(-i\omega_c t) + c.c. \tag{2}$$

The density matrix equations of motion can be driven from the Liouville equation as follows

$$\dot{\rho}_{22} = -2\gamma_1\rho_{22} + i\Omega_p e^{-i\delta t}\rho_{12} - i\Omega_p e^{i\delta t}\rho_{21} - \eta\sqrt{\gamma_2\gamma_3}(\rho_{32} + \rho_{23}),$$

$$\dot{\rho}_{33} = -2\gamma_2\rho_{33} + i\Omega_c(\rho_{13} - \rho_{31}) - \eta\sqrt{\gamma_2\gamma_3}(\rho_{32} + \rho_{23}),$$

$$\dot{\rho}_{12} = [-\gamma_1 + i(\Delta_p + \delta)]\rho_{12} + i\Omega_c\rho_{32} + i\Omega_p e^{i\delta t}(\rho_{22} - \rho_{11}) - \eta\sqrt{\gamma_2\gamma_3}\rho_{13},$$

$$\dot{\rho}_{13} = (-\gamma_2 + i\Delta_c)\rho_{13} + i\Omega_p e^{i\delta t}\rho_{23} + i\Omega_c(\rho_{33} - \rho_{11}) - \eta\sqrt{\gamma_2\gamma_3}\rho_{12},$$

$$\dot{\rho}_{32} = -(\gamma_1 + \gamma_2)\rho_{32} + i(\Delta_p - \Delta_c + \delta)\rho_{32} + i\Omega_c\rho_{12} - i\Omega_p e^{i\delta t}\rho_{31} - \eta\sqrt{\gamma_2\gamma_3}(\rho_{22} + \rho_{33}),$$

$$\rho_{11} + \rho_{22} + \rho_{33} = 1. \tag{3}$$

where the probe-coupling detuning is denoted by $\delta = \omega_1 - \omega_2$. Since the upper states are considered degenerate we assume $\omega_{31} \cong \omega_{21}$, so in consequence $\delta = \Delta_p - \Delta_c$. It is assumed that the two excited states are coupled to the lower level with a common vacuum mode, so the vacuum induced coherence generate interference in the spontaneous emission of the excited levels. Such effect is included in equation (3) via the terms corresponding to $\eta\sqrt{\gamma_2\gamma_3}$ where $\eta = \vec{d}_{12}.\vec{d}_{13}/(|\vec{d}_{12}||\vec{d}_{13}|)$.

In two-photon resonance condition, i.e. $\delta = \Delta_p = \Delta_c = 0$, and for a weak probe field the simple analytical relation for the probe coherence $\rho_{12}$ is given by

$$\rho_{12} = \frac{-2\eta\Omega_c + 8\eta^3\Omega_c - 4\eta^5\Omega_c + \eta\Omega_c^5}{D} + \frac{2\eta^4\Omega_c^2 - 4\eta^2\Omega_c^2 - \eta^2\Omega_c^4}{D}\Omega_p + \frac{4(\eta^2-1)^2 + 2\Omega_c^2(3-3\eta^2+\eta^4) + \Omega_c^4(2-\eta^2)}{D}\Omega_p^* \quad (4)$$

where

$$D = -4 + 12\eta^2(1-\eta^2) + 4\eta^6 + \Omega_c^2(-12+14\eta^2+4\eta^4) + \Omega_c^4(-9+\eta^2) - 2\Omega_c^6.$$

The first term in equation (4) shows the scattering of the coupling field into the probe field via vacuum induced coherence which it depends on the relative phase of the applied field. These terms can be generated even in the absence of the probe field. The second term, proportional to $\Omega_p$, is the direct response of the medium to the probe field which does not depend on the relative phase. The third one proportional to $\Omega_p^*$, shows a counter-rotating contribution. To interpretation the physical processes, we are going to describe the quantum paths involved in the creation of the probe coherence $\rho_{12}$. In Fig. 2, we show the different paths to generate the probe coherence in which the figures 2(a)-2(d) show the paths to establish the different terms in numerator of first term of equation (4). Figures 2(e)-2(g) are related to various terms in numerator of the second term of equation (4). We display the corresponding transitions in table (1). Note that the vacuum induced coherence has a major role in establishing the all contributions of the transition paths.

**Beyond two-photon resonance condition**

The density matrix equations of motion for the mentioned system are solved numerically beyond multi-photon resonance condition using the Floquet frame. "The Floquet theorem and the time-independent Floquet Hamiltonian method are powerful theoretical framework for the study of

bound–bound multiphoton transitions driven by periodically time-dependent fields" [49]. In equations (3), we define a vector $\tilde{R}$ involving all density matrix elements,

Table 1. The transitions path of Fig. 2(a)-2(g).

| Fig. 2(a) | $\eta\Omega_c$ | $\|1\rangle \xrightarrow{\Omega_c} \|3\rangle \xrightarrow{\eta} \|2\rangle$ |
|---|---|---|
| Fig. 2(b) | $\eta^3\Omega_c$ | $\|1\rangle \xrightarrow{\Omega_c} \|3\rangle \xrightarrow{\eta} \|2\rangle \xrightarrow{\eta} \|3\rangle \xrightarrow{\eta} \|2\rangle$ |
| Fig. 2(c) | $\eta^5\Omega_c$ | $\|1\rangle \xrightarrow{\Omega_c} \|3\rangle \xrightarrow{\eta} \|2\rangle \xrightarrow{\eta} \|3\rangle \xrightarrow{\eta} \|2\rangle \xrightarrow{\eta} \|3\rangle \xrightarrow{\eta} \|2\rangle$ |
| Fig. 2(d) | $\eta\Omega_c^5$ | $\|1\rangle \xrightarrow{\Omega_c} \|3\rangle \xrightarrow{\Omega_c^*} \|1\rangle \xrightarrow{\Omega_c} \|3\rangle \xrightarrow{\Omega_c^*} \|1\rangle \xrightarrow{\Omega_c} \|3\rangle \xrightarrow{\eta} \|2\rangle$ |
| Fig. 2(e) | $\eta^2\Omega_c^2\Omega_p$ | $\|1\rangle \xrightarrow{\Omega_c} \|3\rangle \xrightarrow{\Omega_c^*} \|1\rangle \xrightarrow{\Omega_p} \|2\rangle \xrightarrow{\eta} \|3\rangle \xrightarrow{\eta} \|2\rangle$ |
| Fig. 2(f) | $\eta^4\Omega_c^2\Omega_p$ | $\|1\rangle \xrightarrow{\Omega_c} \|3\rangle \xrightarrow{\Omega_c^*} \|1\rangle \xrightarrow{\Omega_p} \|2\rangle \xrightarrow{\eta} \|3\rangle \xrightarrow{\eta} \|2\rangle \xrightarrow{\eta} \|3\rangle \xrightarrow{\eta} \|2\rangle$ |
| Fig. 2(g) | $\eta^2\Omega_c^4\Omega_p$ | $\|1\rangle \xrightarrow{\Omega_c} \|3\rangle \xrightarrow{\Omega_c^*} \|1\rangle \xrightarrow{\Omega_c} \|3\rangle \xrightarrow{\Omega_c^*} \|1\rangle \xrightarrow{\Omega_p} \|2\rangle \xrightarrow{\eta} \|3\rangle \xrightarrow{\eta} \|2\rangle$ |

$$R = (\rho_{11}, \rho_{12}, \rho_{13}, \rho_{21}, \rho_{22}, \rho_{23}, \rho_{31}, \rho_{32})^T \tag{4}$$

The equations of motion (3) can be rewritten compactly in matrix form as follows:

$$\frac{\partial R}{\partial t} + \Sigma = MR, \tag{5}$$

where

$$\Sigma = (-2\gamma_2, 0, -i\Omega_c, 0, 0, 0, i\Omega_c, \eta\sqrt{\gamma_1\gamma_2})^T.$$

Note that $\rho_{33}$ is eliminated because of trace condition $Tr(\rho) = 1$ which is the reason for appearance of the constant term $\Sigma$ in equation (5). [50].

Both the Matrix $M$ and the $\Sigma$ can be separated into the terms with different time dependencies as written below [51].

$$\Sigma = \Sigma_0 + \Omega_P \Sigma_1 e^{-i\delta t} + \Omega_P^* \Sigma_{-1} e^{i\delta t}, \tag{6}$$

$$M = M_0 + \Omega_P M_1 e^{-i\delta t} + \Omega_P^* M_{-1} e^{i\delta t}, \tag{7}$$

where $M_k$ and $\Sigma_k$ ($k \in \{-1, 0, 1\}$) are time-independent. By substituting the relations (6) and (7) into equation (5) we have

$$\dot{R}+\Sigma_0+\Omega_p\Sigma_1 e^{-i\delta t}+\Omega_p^*\Sigma_{-1}e^{i\delta t}=(M_0+\Omega_p M_1 e^{-i\delta t}+\Omega_p^* M_{-1}e^{i\delta t})R, \qquad (8)$$

According to Floquet decomposition [43] which is quite helpful to understand the higher order nonlinearity effects in optical medium. The stationary solution $R$ to Eq. (8) will have only terms at the harmonics of the detuning $\delta$ [46]. By expanding $R$ to the first order probe field Rabi frequency we obtain:

$$R=R_0+\Omega_p e^{-i\delta t}R_1+\Omega_p^* e^{i\delta t}R_{-1} \qquad (9)$$

The coefficients can be driven from equations (5)-(9) as follows

$$R_0=M_0^{-1}\Sigma_0 \qquad (10)$$

$$R_1=(M_0+i\delta)^{-1}(\Sigma_{+1}-M_1 R_0) \qquad (11)$$

$$R_{-1}=(M_0-i\delta)^{-1}(\Sigma_{-1}-M_{-1}R_0) \qquad (12)$$

By using this method we could gain an idea of how sensitive DEM must be the portion effect of probe field and coupling field. Moreover we predict that the oscillatory behavior of DEM beyond two-photon resonance condition can be related to the scattering of the coupling field into the probe field mode at a frequency different from the probe field frequency [42].

On the other hand the reduced entropy provides one tool which can be used to quantify entanglement, although other entanglement measures exist [52]. Mathematically, the bipartite quantum system is called entangled, when its density operator cannot be written as a simple product of the density operator of subsystem, i.e. [5].

$$\rho_{AB}\ne\rho_A\otimes\rho_B. \qquad (13)$$

The atomic system and the vacuum fields are initially in a disentangled pure state which means ($\rho_{11}=1$) and the excited states are empty of population. Reduced entropy of each subsystem will be changed all the while of the interaction between light and atomic ensemble. If the overall system is pure, the entropy of one subsystem can be used to measure its degree of entanglement with the other subsystems. For bipartite pure states, the von Neumann entropy of reduced states is a good measure of entanglement [53, 54]. The quantum mechanical entropy can be defined as:

$$S(\rho) = -Tr\rho \ln \rho \tag{14}$$

which is associated with $\rho$.

The reduced density operator of the atoms as a first subsystem, $(A)$, is defined as:

$$\rho_A = tr_B\{\rho_{AB}\} \tag{15}$$

The reduced density operator of the spontaneous emission, $(B)$, is described by:

$$\rho_B = tr_A\{\rho_{AB}\}, \tag{16}$$

where $\rho_{AB}$ is density operator of pure state for two subsystems $A, B$. Therefore the von Neumann entropy of entanglement is derived as [4]:

$$S(\rho_A) := S(\rho_B) := -tr(\rho_A \ln \rho_A) = -tr(\rho_B \ln \rho_B). \tag{17}$$

For atom-photon disentanglement, the reduced entropy should be equal to zero, and for maximally entangled states for $N$ state particle is $\log_e N$ [4].

Also the triangle inequality relation which it has been presented by Araki and Lieb [53-55] should be satisfied:

$$\left| S^A - S^B \right| \leq S^{AB} \leq S^A + S^B \tag{18}$$

In fact the generated light is entangled when it pass through aforementioned atomic system.

**Results and discussions**

The density matrix equations of motion in a three-level V- type closed- loop atomic system are calculated numerically in two-photon resonance condition and beyond it. For simplicity, it is supposed that $\hbar = 1$. Let us explain our results by interpreting the physical meaning of each plot. All of the parameters in computer's codes are reduced to dimensionless units through scaling by $\gamma_1 = \gamma_2 = \gamma = 1$, and all plots are sketched in the unit of $\gamma$. We are interested in investigating the dynamical behavior of DEM beyond the two-photon resonance condition (TPR).

Figure 3 shows the dynamical behavior of DEM beyond the TPR condition for different parameters of quantum interference $\eta = 0.0$ (solid), 0.5 (dashed) 0.99 (dotted). Other selected parameters are $\gamma_1 = \gamma_2 = \gamma = 1$, $\Delta_c = 10\gamma$, $\Delta_p = 0.0\gamma$, $\delta = -10\gamma$, $\Omega_c = 3\gamma$, $\Omega_p = 0.1\gamma$. It is shown that, the steady state solution can be achieved for the time evolution of the atom-photon entanglement in such a system in multi –photon resonance condition [38]. Here as it can be seen in Fig. 3, DEM has oscillatory behavior versus the time and the stationary solution cannot occur anymore. We predict this behavior can be related to the scattering of the coupling field in probe field mode at a frequency different than the probe frequency.

For understanding the physics behind of this behavior we increase the intensity of coupling field and dynamical behavior of DEM beyond the two-photon resonance condition for $\Omega_c = 20\gamma$, are shown in Fig. 4 for different values of quantum interference parameters $\eta = 0.0$ (solid), 0.5 (dashed) 0.99 (dotted). Other parameters are same as in Fig. 3. It is obvious that, DEM increases by increasing the Rabi frequency of coupling field. Moreover, the amplitude of oscillation becomes smaller in comparing with the total DEM. Our calculation shows that the contribution of the direct response of the medium to the probe field in DEM is negligible in comparing with the contribution of the scattering of the coupling field in the probe field.

The steady state behavior of DEM versus Rabi frequency $\Omega_c$ and quantum interference $\eta$ in the TPR condition is shown in Fig. 5. Used parameters are $\delta = 0.0\gamma$, $\Delta_p = 0.0\gamma$, $\Delta_c = 0.0\gamma$, $\eta = 0.5$. Other parameters are same as in Fig. 3. It is clearly seen that, DEM can be controlled by either the intensity of coupling field or quantum interference. It is realized that, DEM is much more sensitive towards the Rabi frequency of coupling field in comparison with the quantum interference. The effect of the Rabi frequency of the probe and coupling fields on the DEM is shown in Fig. 6 in the TPR condition. Used parameters are same as in Fig. 5. As it is illustrated in Fig. 6, for getting higher value for DEM, the intensity of driving fields should be increased.

Let us turn our attention to the behavior of DEM beyond the TPR condition. In Fig. 7 we display the couple beam portion of DEM behavior versus the detuning of the probe field beyond the TPR condition in different values of quantum interference parameter $\eta = 0.0$ (solid), 0.5 (dashed) 0.99 (dotted). The selected parameters are $\Delta_c = 0.0\gamma$, $\delta = \Delta_p$. Other parameters are same as in Fig. 3. It is shown that, the maximum DEM is obtained at TPR condition and it grows by increasing the quantum interference parameter.

In Fig. 8 by using Floquet decomposition we have chosen the portion of couple beam light to investigate the DEM behavior versus the detuning of coupling field beyond the two-photon resonance condition in different values of quantum interference $\eta = 0.99$ (dashed), 0.5 (solid) 0.0 (dotted). Other parameters are same as in Fig. 3. In fact DEM can be controlled by intensity of the coupling field. The DEM peak located at the $\Delta_c = 0.0\gamma$ is established due to the two-photon transition via a counter-rotating contribution of Eq. (4). In the absence of the vacuum induced coherence, only the phase conjugation of the probe field contributes in preparing of the probe coherence. The second peak located at $\Delta_c = 10.0\gamma$ is due to the transition $|2\rangle - |3\rangle$ via vacuum induced coherence. It is obvious that, in the absence of quantum interference by increasing the detuning of coupling field the disentanglement can be happened. Our investigations showed that, the detuning of the coupling field is important controllable parameter for DEM. Note that the scattering of the coupling field into the probe field via vacuum induced coherence (Fig. 2(a-d)) has a major contribution in establishing of the entanglement in the weak probe field approximation.

**Conclusion**

We studied the atom- photon entanglement in a three- level V-type closed-loop atomic system beyond the two- photon resonance condition. The dynamical behavior of atom- photon entanglement between the dressed atom and its spontaneous emission is studied in semi classical approach beyond the two-photon resonance condition in such a system. The quantum entropy of these two subsystems is investigated by using the von Neumann entropy. It is shown that, the degree of entanglement measure (DEM) can be controlled by the intensity and the detuning of coupling optical field and quantum interference induced by spontaneous emission. Furthermore In the absence of quantum interference for special parameters of Rabi frequency and detuning of driving laser field disentanglement can be occurred. Also the electromagnetically induced transparency condition can be obtained when the system is disentangled.

**Figure Caption**

**Figure 1.** Schematic diagram of the three-level V-type atomic system. The system is driven by two optical laser fields, the laser field which is coupled on the transition $|1\rangle - |2\rangle$ is the probe (weak) field. The spontaneous decays are denoted by the wiggly red lines.

**Figure 2.** Schematic diagram for showing the different paths to generate the probe coherence in the system.

**Figure 3.** Time evolution of DEM beyond the two- photon resonance condition for different values of quantum interference parameters $\eta = 0.0$ (solid), $0.5$ (dashed) $0.99$ (dotted). The other selected parameters are $\gamma_1 = \gamma_2 = \gamma = 1$, $\Delta_c = 10\gamma$, $\Delta_p = 0.0\gamma$, $\delta = -10\gamma$, $\Omega_c = 3\gamma$, $\Omega_p = 0.1\gamma$.

**Figure 4.** Dynamical behavior of DEM beyond the multiphoton resonance condition for $\Omega_C = 20\gamma$ for different parameters of quantum interference $\eta = 0.0$ (solid), $0.5$ (dashed) $0.99$ (dotted). Other parameters are same as in Fig. 3.

**Figure 5.** Steady state behavior of DEM versus Rabi frequency $\Omega_c$ and quantum interference $\eta$. The using parameters are $\gamma_1 = \gamma_2 = \gamma = 1$, $\Omega_p = 0.1\gamma$, $\delta = 0.0\gamma$, $\Delta_p = 3.0\gamma$, $\Delta_c = 3.0\gamma$.

**Figure 6.** Steady state behavior of DEM versus Rabi frequencies of driving fields $\Omega_c$ and $\Omega_p$. Used parameters are $\gamma_1 = \gamma_2 = \gamma = 1$, $\delta = 0.0\gamma$, $\Delta_p = 0.0\gamma$, $\Delta_c = 0.0\gamma$, $\eta = 0.5$.

**Figure 7.** Couple beam portion of DEM behavior versus the detuning of probe field beyond the multiphoton resonance condition in different values of quantum interference $\eta = 0.0$ (solid), $0.5$ (dashed) $0.99$ (dotted), $\Delta_c = 0.0\gamma$ and the other parameters are same as in Fig. 3.

**Figure 8.** The portion of couple beam light for DEM behavior versus the detuning of coupling field beyond the multiphoton resonance condition in different values of quantum interference $\eta = 0.0$ (solid), $0.5$ (dashed) $0.99$ (dotted), $\Delta_p = 0.0\gamma$, and the other parameters are same as in Fig. 3.

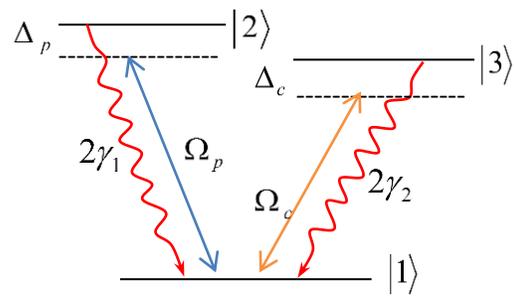

Figure 1.

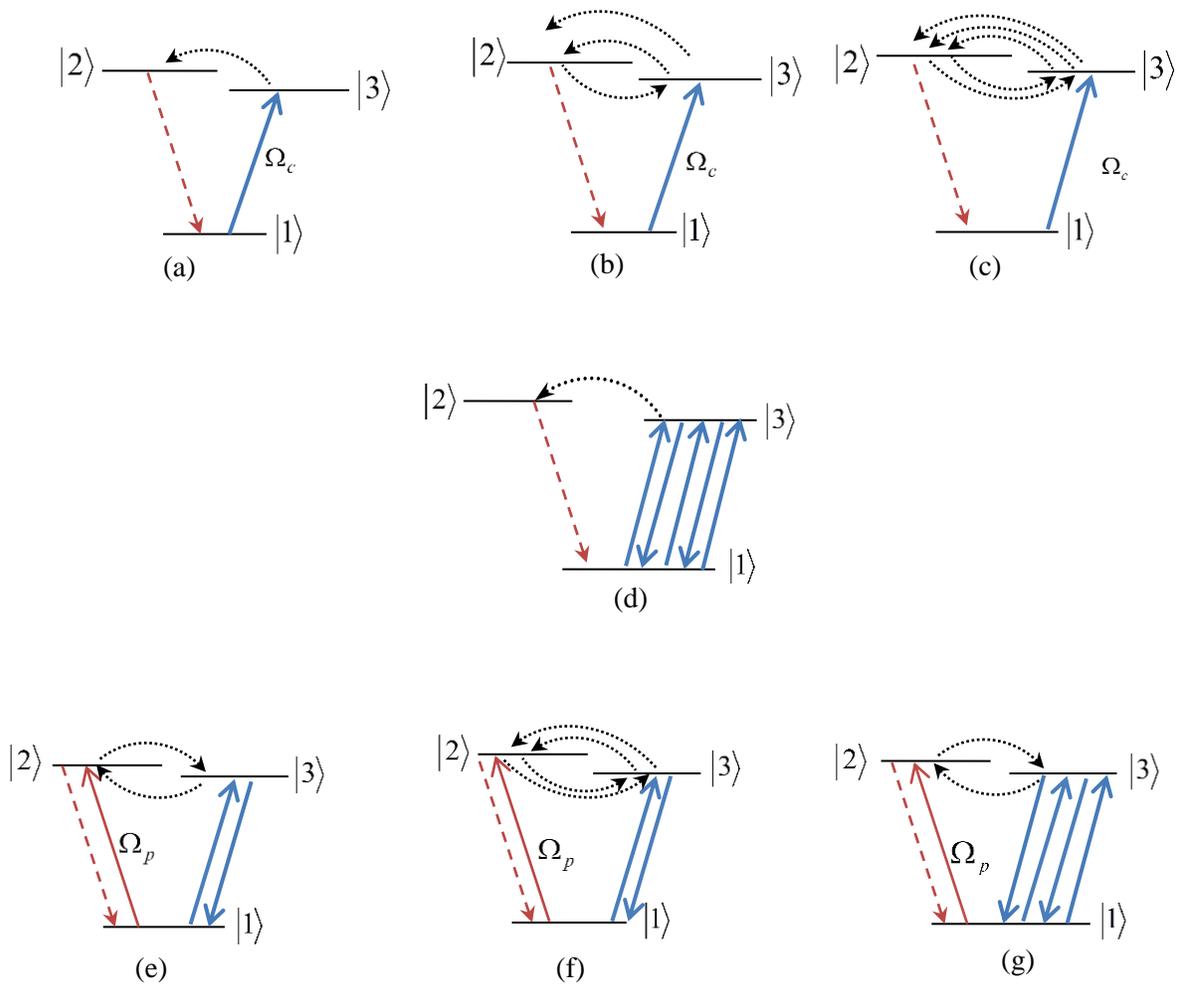

Figure 2

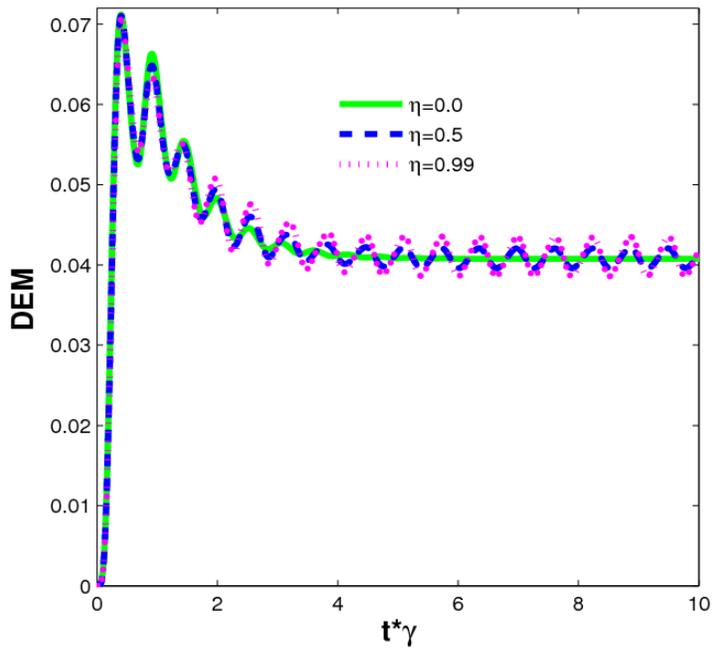

Figure 3.

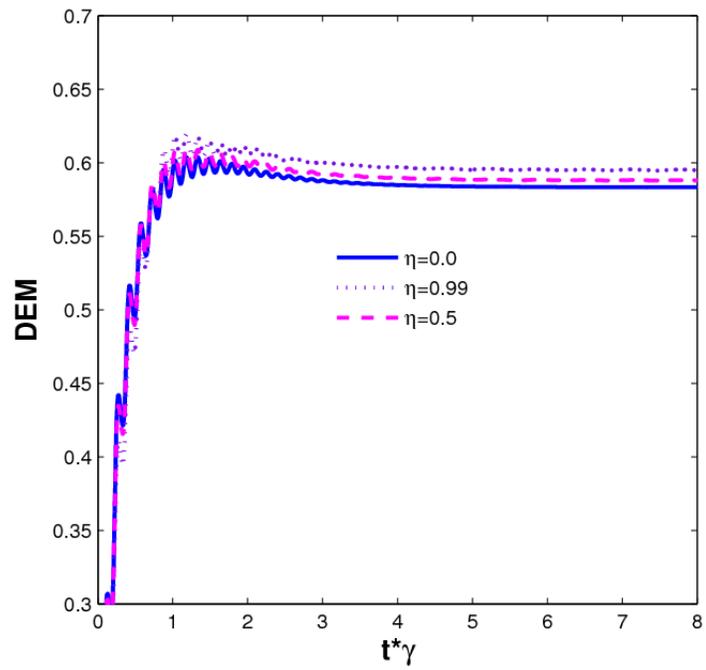

Figure 4.

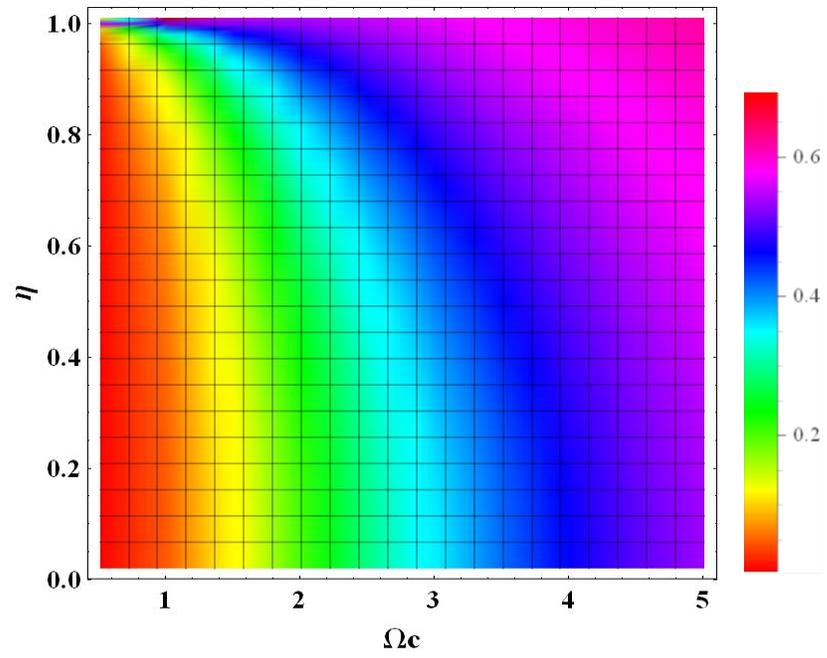

Figure 5.

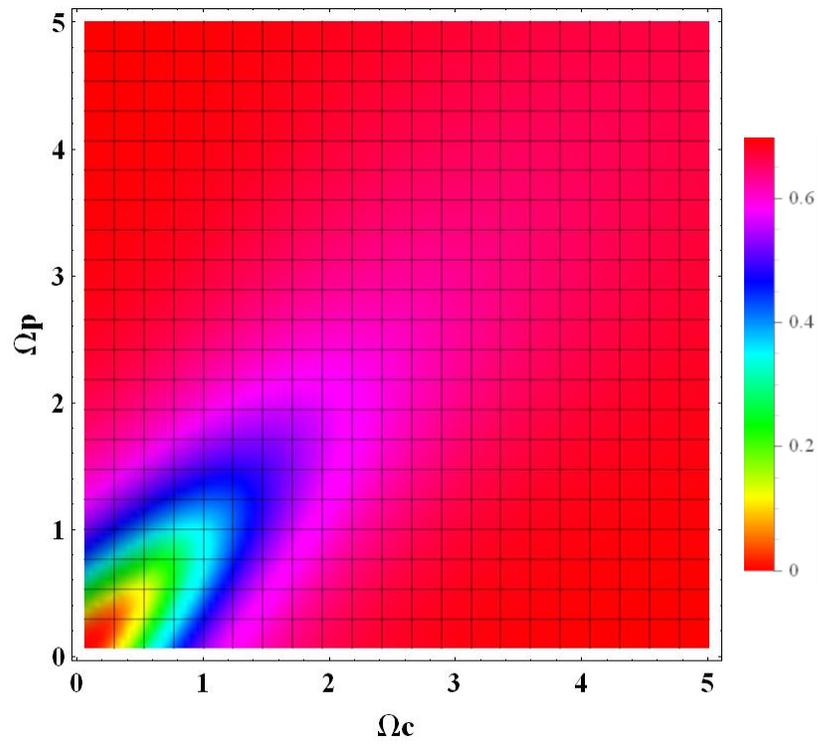

Figure 6.

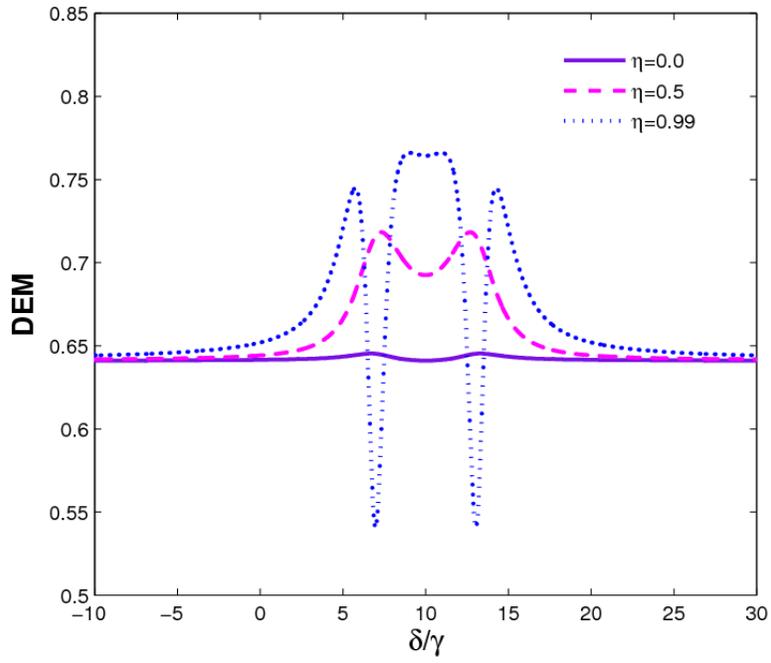

Figure 7.

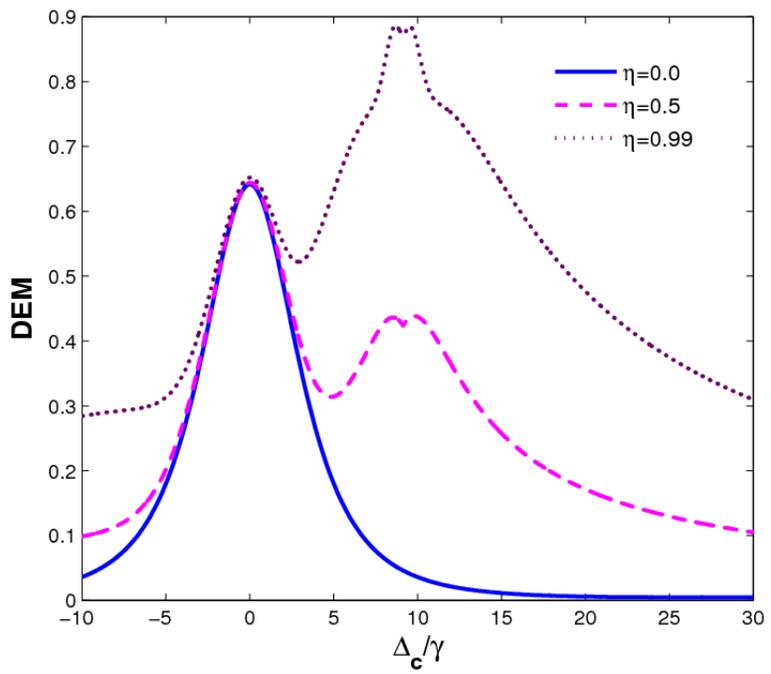

Figure 8.